\documentclass[twocolumn]{aastex63} \usepackage{amsmath} \usepackage{booktabs}
\usepackage[utf8]{inputenc} \usepackage{xcolor} \usepackage{graphicx} \usepackage{hyperref} \usepackage[caption=false]{subfig}
\usepackage{wrapfig}

\begin{document}

\title{Realistic observing scenarios for the next decade of early warning detection of binary neutron stars}

\author[0000-0001-9769-531X]{Ryan Magee}
\affiliation{LIGO, California Institute of Technology, Pasadena, CA 91125, USA}

\author{Ssohrab Borhanian}
\affiliation{Theoretisch-Physikalisches Institut, Friedrich-Schiller-Universit{\"a}t Jena, D-07743 Jena, Germany}

\begin{abstract}

We describe realistic observing scenarios for early warning detection of binary
neutron star mergers with the current generation of ground-based
gravitational-wave detectors as these approach design sensitivity. Using Fisher
analysis, we estimate that Advanced LIGO and Advanced Virgo will detect
one signal before merger in their fourth observing run provided they
maintain a 70\% duty cycle. 60\% of all observations and
8\% of those detectable 20 seconds before merger will be localized to
$\lesssim 100 \thinspace \mathrm{deg}^2$. If KAGRA is able to achieve a
25 Mpc horizon, these prospects increase to $\lesssim 2$  early detections with
70\% of all BNS localized to $\lesssim 100 \thinspace
\mathrm{deg}^2$ by merger.  As the AHKLV network approaches design sensitivity
over the next $\sim10$ years, we expect up to 1 (14) detections made
100 (10) seconds before merger. Although adding detectors to the HLV
network impacts the detection rate at $\lesssim 50\%$ level, it improves
localization prospects and increases the completeness of compact binary
surveys. Given uncertainties in sensitivities, participating detectors, and
duty cycles, we 
%include a data release that allows for full generalizability of 
consider 103 future detector configurations so electromagnetic
observers can tailor preparations towards their preferred models. 

\end{abstract}

\section{Introduction} \label{sec:intro}

Advanced LIGO~\citep{TheLIGOScientific:2014jea}, Advanced
Virgo~\citep{TheVirgo:2014hva}, and KAGRA~\citep{Akutsu:2018axf} have enjoyed
remarkable success since the first detection of gravitational waves (GWs) from a
binary black hole merger in 2015~\citep{LIGOScientific:2016aoc}.  Since
then, analyses by the LIGO-Virgo-KAGRA collaboration (LVK) have uncovered a
growing population of binary black holes, binary neutron stars (BNSs), and
neutron star - black hole binaries
(NSBHs)~\citep{LIGOScientific:2018mvr,LIGOScientific:2020ibl,LIGOScientific:2021usb,LIGOScientific:2021djp}.
Analyses of public data~\citep{Trovato:2019liz} have confirmed many of these
detections and hinted at other promising candidates lurking deeper in the
noise~\citep{Nitz:2018imz,Magee:2019vmb,Zackay:2019tzo,Venumadhav:2019lyq,Nitz:2020oeq,Nitz:2021uxj,Olsen:2022pin}.

GW observations coincident with other astrophysical signals such as
electromagnetic radiation or particles are a highly sought-after subclass of
so-called \emph{multi-messenger} detections.
Even before the first detection of GWs, various studies investigated what to expect from electromagnetic follow-up efforts during the Advanced LIGO
and Advanced Virgo era~\citep{Cannon:2011vi, Singer:2014qca}. The first
multi-messenger detection with GWs,
GW170817~\citep{LIGOScientific:2017vwq,LIGOScientific:2017ync}, was
serendipitous in nature, and led to an explosion in works focused on
facilitating additional discoveries.

In particular, there has been an increasing focus on \emph{early warning} (or
pre-merger) detection and localization of BNSs and
NSBHs~\citep{Sachdev:2020lfd,Nitz:2020vym,Singh:2020lwx,Yu:2021vvm,Tsutsui:2021izf,Kovalam:2021bgg}.
Several of these works, in
particular~\citep{Sachdev:2020lfd,Nitz:2020vym,Singh:2020lwx,Kovalam:2021bgg}, focus on
BNS detection for the current generation of ground-based detectors,
though many optimistically assume a 100\% duty cycle and sensitivities that may
prove difficult to reach~\citep{Akutsu:2018axf,Washimi:2020slk}.
More recently, there has also been a focus on the infrastructure necessary to
realize early warning alerts with an emphasis on latencies expected in the
LVK's fourth observing run (O4)~\citep{Magee:2021xdx}.

In this study, we investigate projected observing scenarios of current
generation ground-based detectors for early warning detection using
well-established Fisher analysis techniques~\citep{PhysRevD.47.2198,PhysRevD.46.5236,Cutler:1994ys,
Poisson:1995ef, Balasubramanian:1995bm}. We extend previous studies of BNSs in
three major ways. First, we estimate the localization for 103
combinations of detectors and detector sensitivities for Advanced LIGO,
Advanced Virgo, and KAGRA in O4 as well as at their projected {\em design}
sensitivity (referred to as O5). Second, we provide the probability density
distributions of the 90\%-credible sky area as a function of early warning
time to enable the consideration of arbitrary network combinations, duty
factors, and early warning detection times. We 
%publicly release the data associated with our simulations~\href{} and we 
compare our results to similar
work previously carried out \citep{Sachdev:2020lfd,Nitz:2020vym,KAGRA:2013rdx}.
Finally, we include the impact that the KAGRA detector would have at three
projected and one realized sensitivity in
light of recent construction difficulties~\citep{Akutsu:2018axf,
Washimi:2020slk}.

\section{Methods}

% emphasize benefits of fisher analyses

Bayesian approaches such as full parameter
estimation~\citep{Veitch:2014wba,Ashton:2018jfp} and the \texttt{bayestar}
code~\citep{Singer:2015ema} are presently used to provide the most accurate
localizations possible for compact binary mergers. Although
full parameter estimation is accurate, it takes $\mathcal{O}(\mathrm{hours}
\, \textsc{-}\, \mathrm{days})$. \texttt{bayestar} does not sample over the masses or spins of the binary and is able to provide comparable localizations in $\mathcal{O}(1
\thinspace \mathrm{second})$ when run in parallel (or $\mathcal{O}(1 \thinspace
\mathrm{minute})$ on a single thread). \cite{Singer:2014qca} showed that the
two methods largely agree with each other.

For large trade studies, Fisher analysis methods are
often favored since they only depend on the
characteristics of the gravitational waveform model and detectors under
investigation. They provide a simple
and fast way of estimating the information a signal contains on the waveform
model's parameter domain.
Here, we use the Fisher analysis-based software package
\texttt{gwbench}~\citep{Borhanian:2020ypi} to benchmark the measurement
capabilities of 103 detector network configurations detailed in the next section. Since
this study is focused on BNS signals which do not merge in the most sensitive bands of
current generation detectors, we consistently employ a simple inspiral waveform
model, TaylorF2~\citep{Sathyaprakash:1991mt, Blanchet:1995ez, Blanchet:2005tk,
Buonanno:2009zt} to estimate the expected measurement uncertainties for the
following parameters:
$\mathcal{M}, \eta, D_L, t_c, \phi_c, \iota, \psi, \alpha, \delta$. These denote
the chirp mass, symmetric mass ratio, luminosity distance, coalescence time,
coalescence phase, inclination, polarization angle, right ascension, and
declination, respectively. Finally, we can estimate the uncertainty in sky
localization via the 90\%-credible sky area following
\cite{Cutler:1997ta} and \cite{Barack:2003fp} as
\begin{equation}
\Delta \Omega_{90} = 2\pi \ln(10)\, |\cos \delta |\, \sqrt{\mathrm{Var}(\alpha)\mathrm{Var}(\delta) - \mathrm{Cov}^2(\alpha,\delta)}
\end{equation}
where $\mathrm{Var}(\alpha)$ and $\mathrm{Var}(\delta)$ are the variances
of the right ascension and declination, respectively, and
$\mathrm{Cov}(\alpha,\delta)$ is the covariance between the right ascencion and
declination. There are, however, several limitations to Fisher analysis
approaches~\citep{Vallisneri:2007ev,Rodriguez:2013mla}. Most well-known is that
they are only
valid in the high signal-to-noise ratio (SNR) limit ($\gtrsim 10$ per detector)~\citep{Vallisneri:2007ev}.
Further, special considerations are necessary while interpreting sky
localizations: Fisher analyses estimate the overall uncertainty and therefore
cannot estimate proximity of regions of probability in the sky. Additionally,
they can exhibit mirror degeneracies~\citep{Fairhurst:2009tc} and scale differently with the SNR than coherent Bayesian approaches~\citep{Berry:2014jja}.
As~\cite{Cannon:2011vi} noted, Fisher analysis estimates tend to be
optimistic. We quantify this bias of our approach in comparison to
\texttt{bayestar} in Section~\ref{sec:comparison}.

We do not impose any single detector SNR thresholds; note that this means low
SNRs in one detector can contribute to the network SNR threshold used as a
detection criterion. In practice, we expect the bias introduced by this effect
to be small. Unless otherwise noted, we assume a network SNR detection threshold
of 15. In all cases, we conservatively assume that no localizations can be provided
when only one detector is operating.

\begin{figure}
\includegraphics[width=\columnwidth]{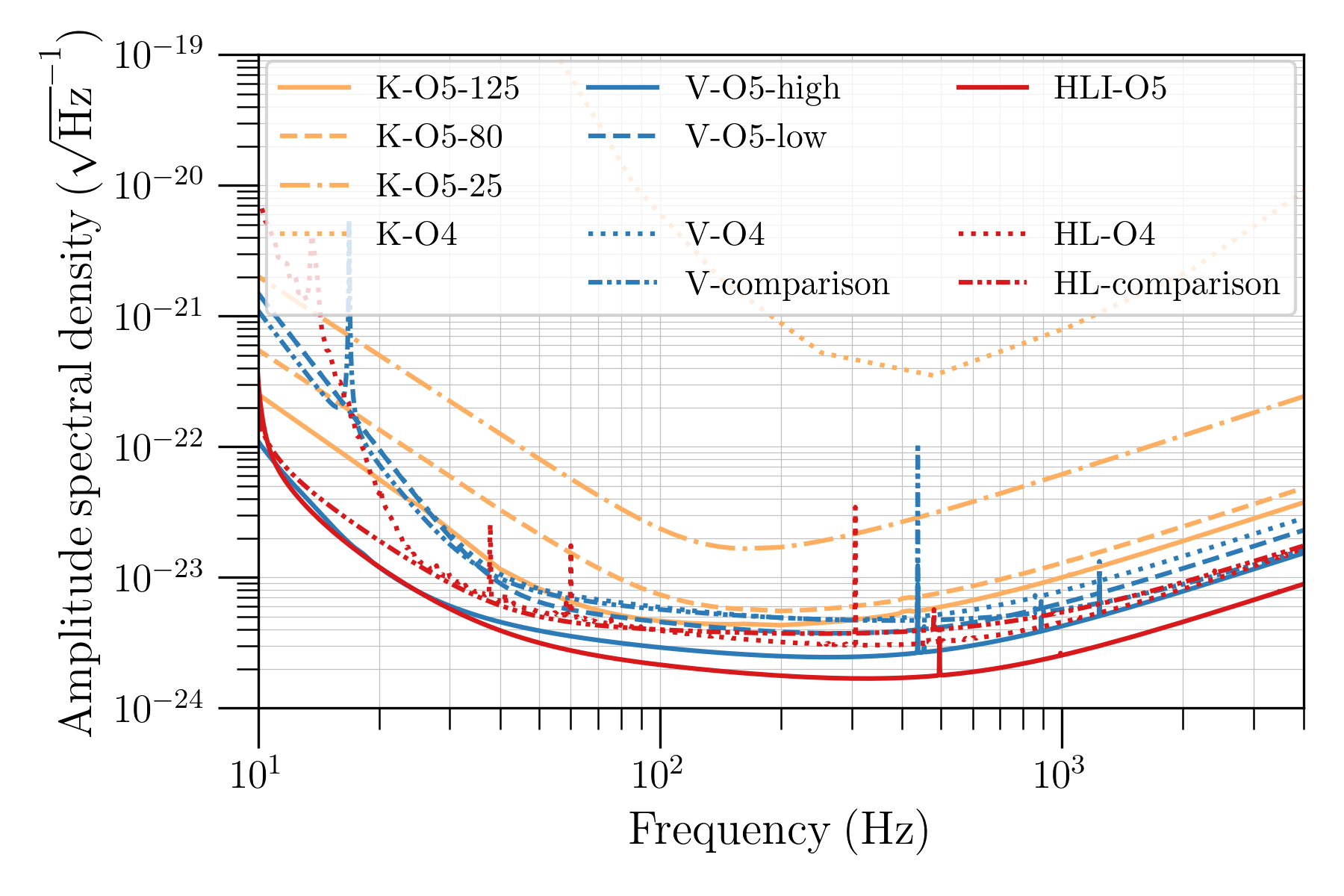} \caption{The
amplitude spectral densities (ASD) assumed for this trade study. With the
exception of the O4 KAGRA ASD that was obtained through digitization of Figure
1 from~\cite{Washimi:2020slk}, all other ASDs are obtained from the observing
scenarios data release. ASDs are labeled as they appear in the data release
and/or in the publicly provided estimates
in~\href{https://dcc.ligo.org/LIGO-T2000012/public}{LIGO-T2000012}.}
\label{fig:psds}
\end{figure}

\begin{figure*}[hbt!]%
\centering
\includegraphics[width=\linewidth]{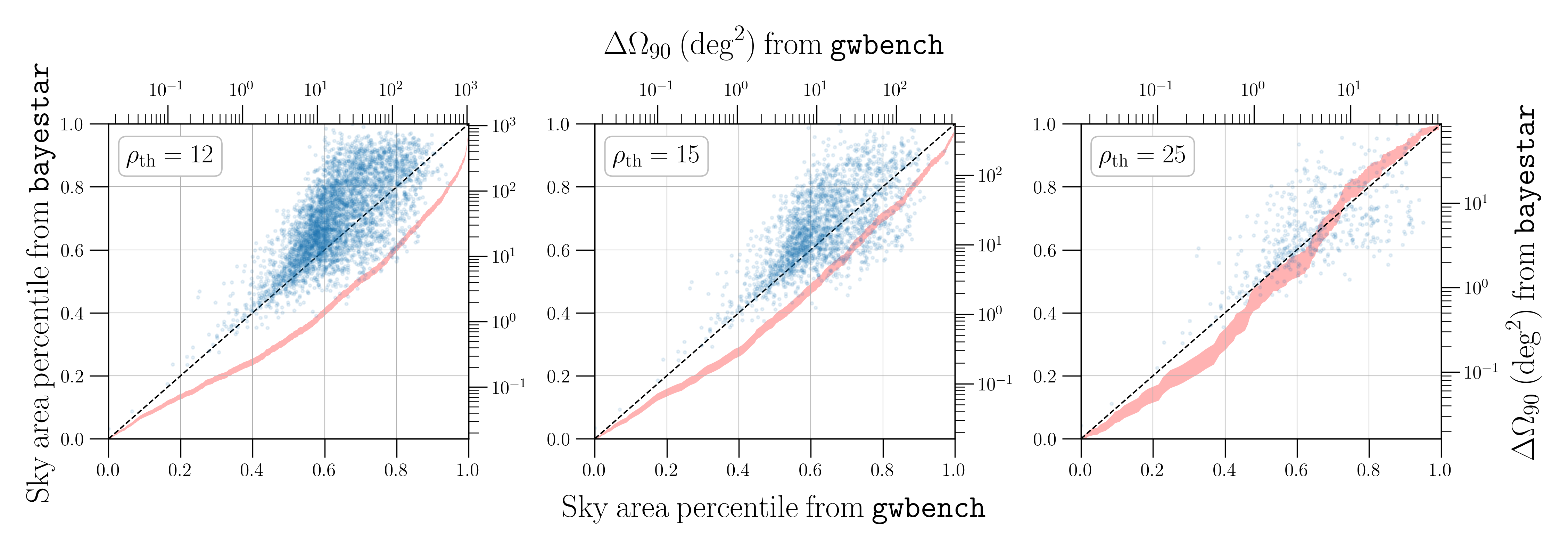}\label{fig:pp-12}
\caption{\label{fig:pp}A comparison of the
90\% confidence interval computed by \texttt{bayestar} vs. the Fisher matrix
formalism used here (\texttt{gwbench}) for events with network SNR thresholds $\rho_{\rm th}$ of 12, 15, and 25. Each point represents a localization measurement obtained by both \texttt{bayestar} and \texttt{gwbench}.
Note that the high SNR (small localization) events largely
agree, but that this agreement becomes statistical in nature for larger areas.
The bias between these two methods is more easily visible in the overlaid
$pp$-plot comparing the 90\%-credible sky area percentiles obtained by
\texttt{gwbench} and \texttt{bayestar}. We include uncertainties on the
percentiles measured due to the size of our population, which is dependent on the SNR threshold used.  We find that \texttt{gwbench} systematically underestimates the size of the confidence interval at SNR 12.}
\end{figure*}

\section{Population and Networks}

\paragraph{Binary neutron star population}
Although two probable NSBH systems were recently observed by Advanced LIGO and
Advanced Virgo~\citep{LIGOScientific:2021qlt}, we restrict the analysis here to
BNS populations due to uncertainties in the NSBH population and poor early
warning and localization prospects for NSBHs. We consider the same
astrophysically motivated source population of BNSs as
in~\citep{Sachdev:2020lfd}. The source-frame component masses are drawn from a
Gaussian distribution with mean mass $1.33 M_\odot$ and standard deviation
$0.09 M_\odot$. The source-frame masses are further limited such that $1.0
M_\odot < m_2 < m_1 < 2.0 M_\odot$. This population is modeled after galactic
observations of BNS~\citep{Ozel:2016oaf}. We note that the masses inferred from
GW190425~\citep{LIGOScientific:2020aai} are in tension with this population.
This could be an indication that galactic measurements are not representative
of the broader population of neutron stars. Results from the LVK's recent
population analysis seems to support this
claim~\citep{LIGOScientific:2021psn}.  We neglect these uncertainties here and
naively apply the most recent BNS merger rate estimates~\citep{LIGOScientific:2021psn} to our population,
though we note that (1) our method can be quickly rerun to produce
estimates for arbitrary populations, (2) that as pointed out
in~\cite{Nitz:2020vym}, our results can be scaled to systems of arbitrary mass,
and (3) that expected BNS localizations do not appear to significantly depend
on the specifics of the population~\citep{Pankow:2019oxl}.

\paragraph{Networks}

We examine 103 GW detector networks for O4
and O5 that arise from 11 different projected sensitivity curves, summarized in
Figure~\ref{fig:psds}. For O4, we consider combinations of
Hanford-KAGRA-Livingston-Virgo (HKLV) networks that contain at least two of the
HLV detectors, with HLV sensitivities as described in the latest observing
scenario\footnote{We use the publicly provided PSDs:
\href{https://dcc.ligo.org/LIGO-T2000012/public}{LIGO-T2000012}.}.
Perhaps the biggest question for O4 is the level to which KAGRA will be able to
participate~\citep{o4lvemupdate}. In this work, we consider two possible KAGRA sensitivities for
O4 with $1 \thinspace \mathrm{Mpc}$ and $25 \thinspace \mathrm{Mpc}$ BNS
detection horizons. We regard this to be a more realistic update to the
recent O4 early warning detection and localization estimates provided
in~\cite{Magee:2021xdx} which assumed a horizon of $80 \thinspace \mathrm{Mpc}$,
especially in light of the recent LVK announcement suggesting KAGRA will start
O4 with $1 \thinspace \mathrm{Mpc}$ horizon~\citep{o4lvemupdate}. The KAGRA
sensitivities were digitized from Figure 1 in~\cite{Washimi:2020slk}.

For O5, we consider a 5-detector network, AHKLV including LIGO-Aundha~\citep{Saleem:2021iwi},
previously LIGO-India~\citep{M1100296}. Following~\cite{KAGRA:2013rdx}, we
assume that Aundha, Hanford, and Livingston are all able to achieve comparable
sensitivities (e.g. the Advanced LIGO design sensitivity). We
compute all network combinations where at least two of the AHLV detectors are operating.
We consider two separate, publicly available sensitivities for Virgo, and
three sensitivities for KAGRA, assuming that by O5 KAGRA will achieve either
$25 \thinspace \mathrm{Mpc}$, $80 \thinspace \mathrm{Mpc}$, or $125 \thinspace
\mathrm{Mpc}$ BNS detection horizons outlined in the observing scenarios
review.

\begin{figure*}[ht!]
\centering
\includegraphics[width=0.8\textwidth]{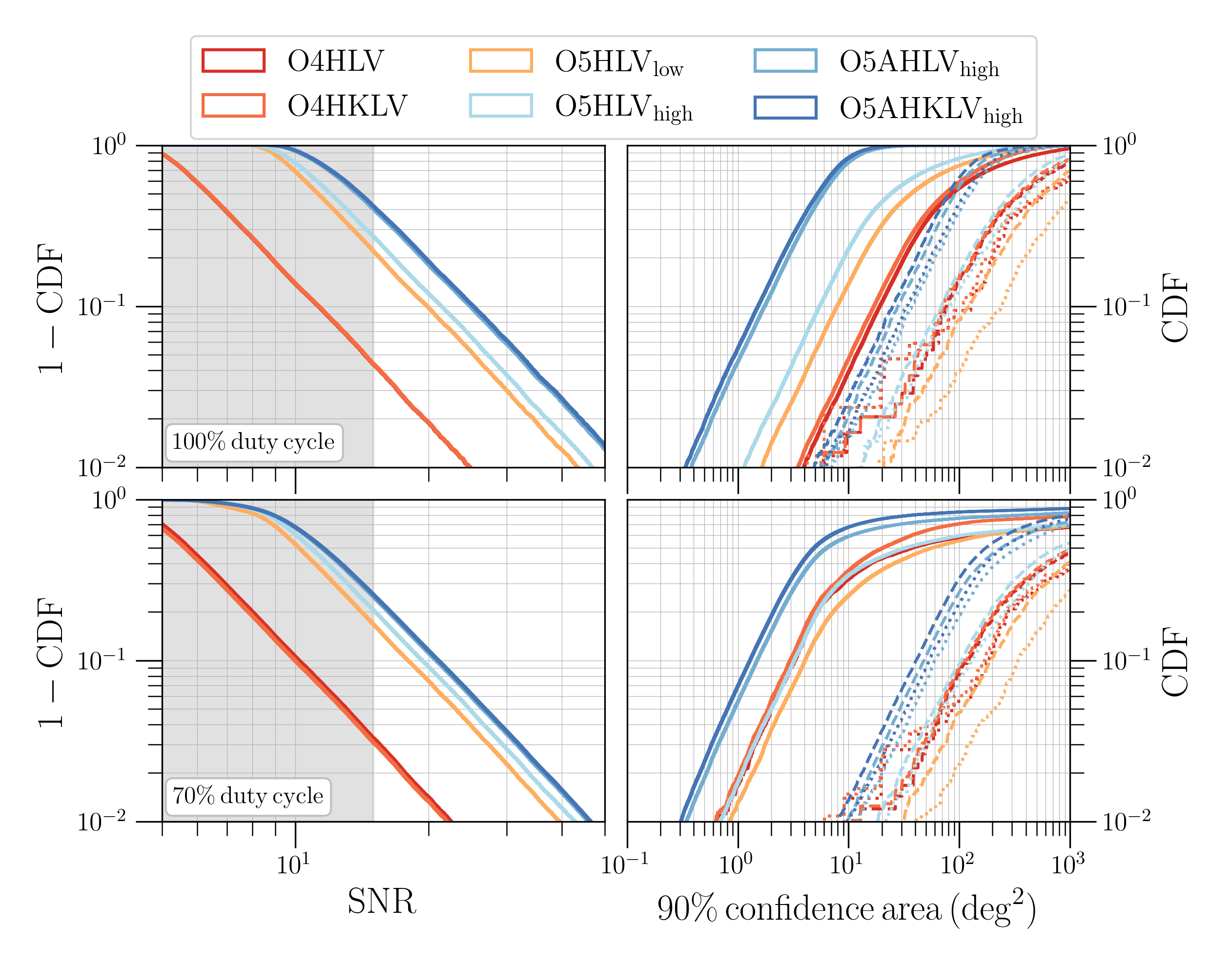}
\caption{Cumulative SNR (left) and localization (right) distributions for six
sample network configurations operating at 100\% (top) and 70\% (bottom) duty
cycles. We include predictions for the early warning localization distributions
obtained  $0\,{\rm s}$ (solid), $20\,{\rm s}$ (dashed), and $40\,{\rm s}$ (dotted) before
merger. The 70\% duty cycle cumulative distributions assume single detector candidates are not
localized (e.g. they are normalized to two or more detector networks).
\label{fig:cdf}} \end{figure*}

\section{Results}
\subsection{Comparison to known localizations}
\label{sec:comparison}
In order to quantify any biases introduced by the Fisher analysis, we compare
our localization estimates from \texttt{gwbench} to those computed by
\texttt{bayestar} for all simulated signals recovered by the full search
presented in~\cite{Sachdev:2020lfd}. Figure~\ref{fig:pp} shows a comparison of
the 90\%-credible sky area computed via each method at three network SNRs $\rho_{\rm net}$ (right and top axes), together
with a $pp$-plot comparing the percentiles associated with each localization
estimator (left and bottom axes). In general, we find that \texttt{gwbench} and
\texttt{bayestar} agree to within a factor of a few depending on the SNR
threshold used. At SNR 12, the 50th (90th) percentiles agree to within
a factor of $\sim3$ ($\sim8$). At SNRs 15 and 25, this improves to $\sim2$ ($\sim6$)
and $< 2$ ($<6$), respectively.
 
Although localizations largely agree on the event-by-event level, we empirically
find that at $\rho_{\rm net}\lesssim 15$ there are significant biases between
the expected localization distributions obtained when compared to
\texttt{bayestar}. The $pp$-plots overlaid in Figure~\ref{fig:pp} show that, in
general, the Fisher analysis systematically underestimates the size of the
90\%-credible sky area. This effect lessens in severity as the SNR threshold is
increased.  For $\rho_{\mathrm{net}} \geq 25$, the bias has mostly disappeared.

We assert that the statistical
agreement between the two methods is trustworthy for
systems with $\rho_{\mathrm{net}} \geq 15$, and accurate to $\lesssim1$ order of
magnitude at lower SNRs. We therefore assume a detection threshold of
$\rho_{\mathrm{net}} \geq 15$ for the remainder of this work and assume the
individual localizations produced are accurate to within a factor of a few.

\subsection{O4}

\begin{figure}
\includegraphics[width=0.5\textwidth]{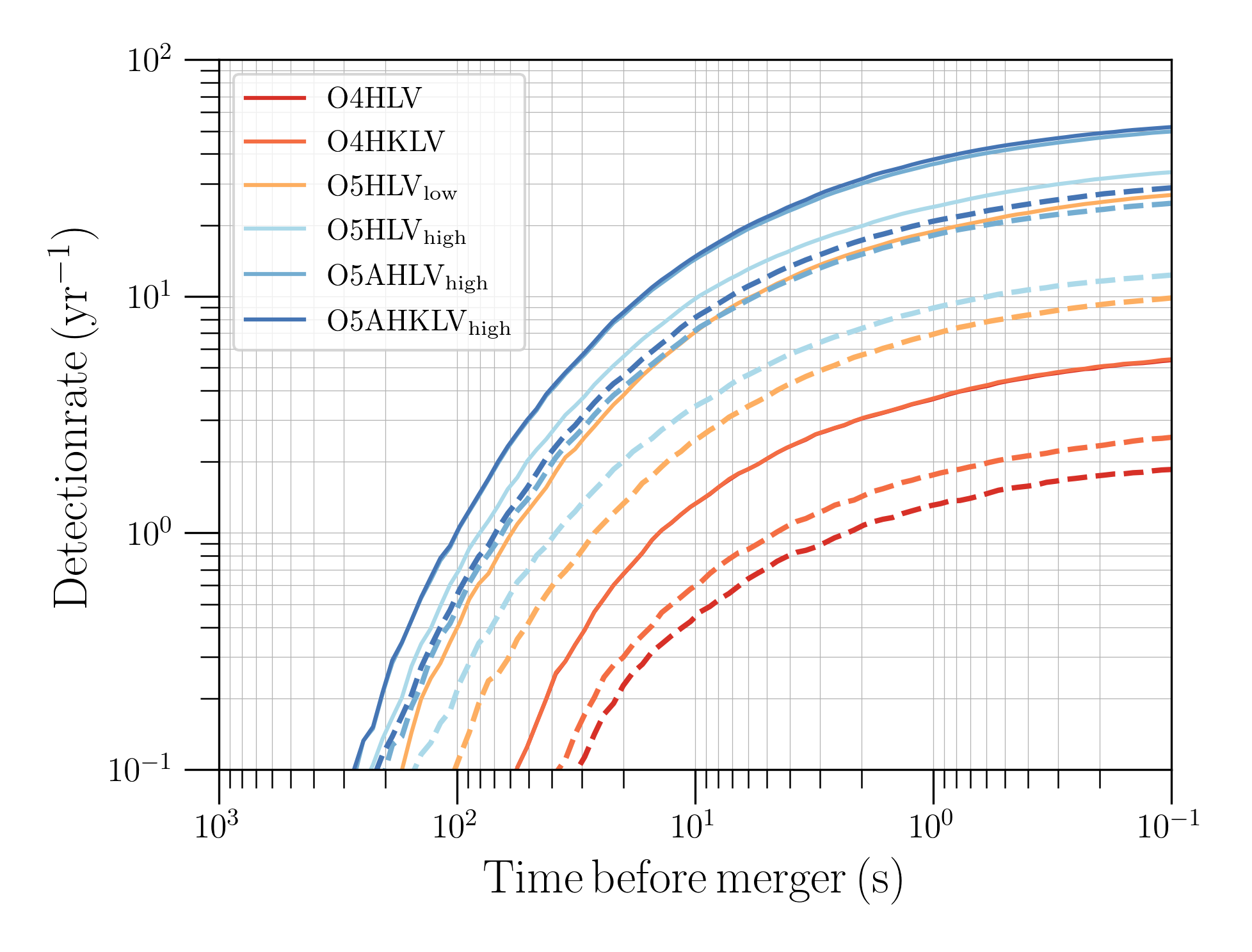}
\caption{The expected number of detections per year vs the time detected before
merger. Latencies associated with data acquisition, transfer, analysis, and
enrichment are not included. The solid (dashed) lines show expectations for
6 different networks operating at 100\% (70\%) duty cycle. The solid
lines of $\textsc{O4HKLVK}$ and $\textsc{O5AHKLV}_{\rm high}$ lie
directly on top of the respective
networks without KAGRA, $\textsc{O4HLV}$ and $\textsc{O5AHLV}_{\rm high}$. All lines
assume a median BNS merger rate of 470 $\mathrm{Mpc}^{-3} \mathrm{yr}^{-1}$. We do not include
uncertainties associated with that measurement in this plot.
\label{fig:detections}}
\end{figure}

Figures~\ref{fig:cdf} and~\ref{fig:detections} present the cumulative SNR and
localization distributions and the expected yearly early warning detection
rates, respectively, for six representative O4 and O5 networks:
$\textsc{O4HLV}$, $\textsc{O4HKLV}$, $\textsc{O5HLV}_{\rm low}$,
$\textsc{O5HLV}_{\rm high}$, $\textsc{O5AHLV}_{\rm high}$, and
$\textsc{O5AHKLV}_{\rm high}$. We highlight two early warning times, $20\,{\rm
s}$ and $40\,{\rm s}$ before merger, which are motivated by data analysis
latencies~\citep{Magee:2021xdx} and approximate $\mathcal{O}(10\, {\rm s})$ slew
times.

It was recently announced that KAGRA is expected to join O4 with a horizon of
at least $\sim 1 \thinspace \mathrm{Mpc}$~\citep{o4lvemupdate}. As expected, we find that there is
no impact on the network sensitivity and negligible impact on the localizations
achieved when this HKLV network is compared to HLV. However, if KAGRA is able to
reach $25 \thinspace \mathrm{Mpc}$, we find an average $\sim 40\%$ reduction
in the 90\%-credible sky areas for events with SNR $\geq 15$, though the
number of expected detections increases at less than the percent level, see
Figure~\ref{fig:cdf}.  Thus
while a moderately sensitive KAGRA in O4 will not increase the number of
detections, it will greatly improve the localization for existing detections.

As shown in Figure~\ref{fig:detections}, early warning detections in O4 are
likely to be exceedingly rare; we expect $\lesssim 1$ detection made to be made early enough to overcome analysis latencies (e.g. $20\,{\rm s}$
early) per year. Of these observations, we expect 10 $\textendash$ 20\% to have
localizations $\lesssim 100 \mathrm{deg}^2$. The presence/absence of KAGRA has
negligible impact on the localization or abundance of early warning detections. 

\subsection{O5 / Design}

For the 3-detector HLV network, we find only a moderate $\lesssim 5\%$
difference in the number of detected events above our SNR threshold as the Virgo
sensitivity is varied (left column Figure~\ref{fig:cdf}), though the
localization distributions noticeably shift. We estimate that up to 20\% (13\%)
of all detected BNS will have 90\%-credible sky areas $\lesssim 10\,{\rm deg^2}$ for
the high (low) sensitivity Virgo projections. We expect
up to 16\% (8\%) of signals detected $20\,{\rm s}$ early and 13\% ($4$\%) of
signals detected $40\,{\rm s}$ early to be localized to $\lesssim 100\,{\rm
deg^2}$.

\begin{figure*}
\begin{interactive}{animation}{movie.mp4}
\includegraphics[width=\textwidth]{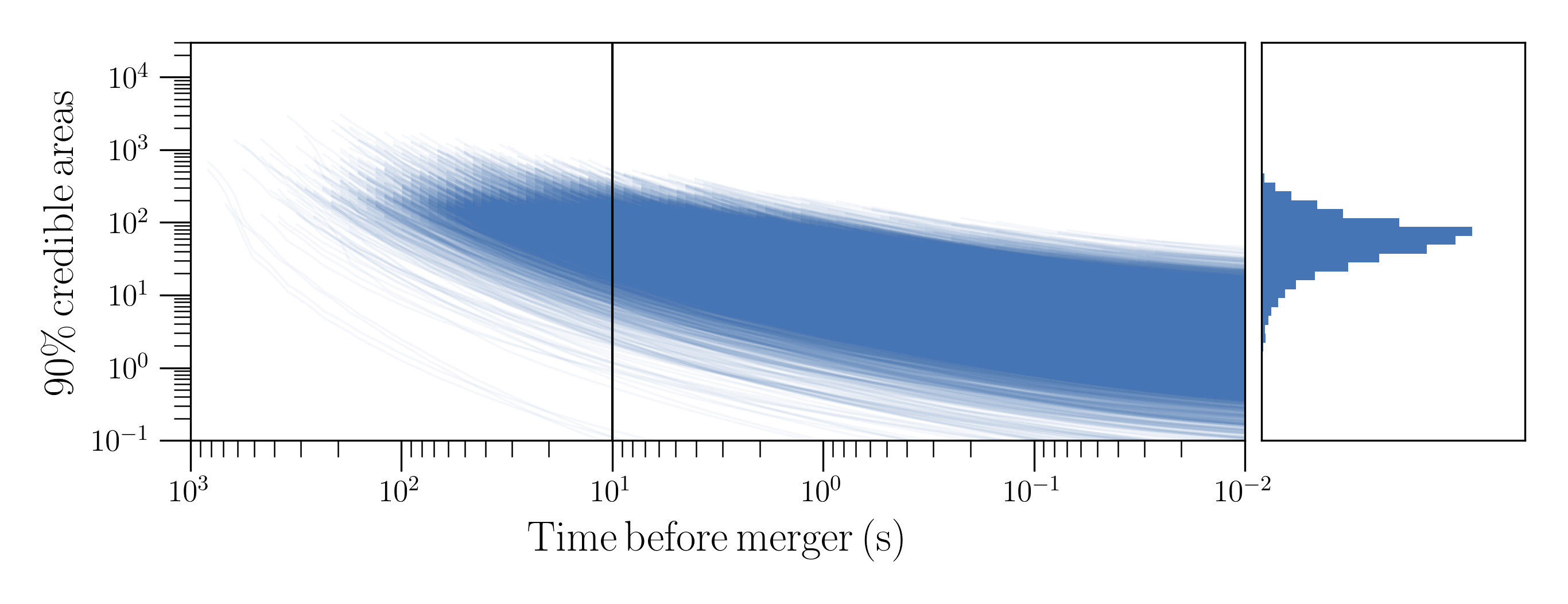}
\end{interactive} \caption{\label{fig:design} Here we show the expected 90\% confidence areas as
a function of detected time before merger for an idealized, design
AHKLV network acting at 100\% duty cycle. Each line tracks one simulation.
The right part shows a histogram of expected localizations at the time
indicated by the vertical black line (e.g. $10\,{\rm s}$ before merger). At 10
s before merger, we expect most events to have localizations
$\mathcal{O}(100 \thinspace \mathrm{deg}^2)$. By merger, this improves to $\mathcal{O}(1 \thinspace
\mathrm{deg}^2)$. Simulations are only tracked once a network SNR $\geq 15$
is reached. An animation of this figure that depicts the localization evolution
is available \href{https://dcc.ligo.org/LIGO-P2200010/public}{online}.}
\end{figure*}

LIGO-Aundha has an even larger impact. Its addition to the HLV network suggests
we expect up to 80\% of detected BNS to be localized to $\lesssim 10\,{\rm
deg^2}$ by merger, dropping to $\lesssim 3$\% and $\lesssim 1$\%, respectively,
$20\,{\rm s}$ and $40\,{\rm s}$ before merger. The addition of KAGRA operating at 125 Mpc has a small
impact on both the detection and localization when compared to the AHLV network
operating at 100\% duty cycle.  For this best case network, we also explicitly compute the evolution of the
90\%-credible sky areas as a function of time before merger in
Figure~\ref{fig:design}, in addition to the two fiducial early warning times,
$20\,{\rm s}$ and $40\,{\rm s}$, in Figure~\ref{fig:cdf}.

\subsection{Duty cycle impact on localizations and detections}

Included in Figure~\ref{fig:cdf} is a comparison of networks operating at 100\%
vs 70\% duty cycle\footnote{Chosen to match the duty cycle in the LVK's
observing scenario document~\citep{LIGOScientific:2021psn}.}.  The real benefit
to networks with a large number of detectors is clear. While at 100\% duty cycle
there is little difference in localizations produced by 4 and 5-detector
networks, there is a large benefit for networks that can only operate at
moderate duty cycles. A 4-detector network at 70\% duty cycle operates with all
4 detectors only 24\% of the time; for a 5-detector network, there are at least 4
detectors active 53\% of the time. The extra detector greatly increases the robustness
of the global detector network. This effect is well demonstrated by the
$\mathrm{O5AHLV}_{\mathrm{high}}$ and $\mathrm{O5AHKLV}_{\mathrm{high}}$ curves
in the right panels of Figure~\ref{fig:cdf}.

The impact on detection is also easily visualized. Figure~\ref{fig:detections}
shows the expected number of detections at a fiducial BNS rate of 470
$\mathrm{Gpc}^{-3}\mathrm{yr}^{-1}$ for networks operating at 100\% and 70\%
duty cycles, respectively. In all scenarios considered, down time decreases the
number of expected detections by a factor of a few.

\section{Outlook and discussion}

Early warning detection will facilitate the capture of prompt, rapidly fading
emission associated with BNS mergers. We find that even the most optimistic
scenarios for O4 predict $\sim1$ BNS detected before merger per year, with
localizations $\gtrsim 100 \, \mathrm{deg}^2$. These detections will likely be
too poorly localized for optical facilities to follow-up. We expect wide-field
observatories such as the Murchinson Wide Field Array \citep{tingay2013} and the
space-based Fermi Gamma-ray Burst Monitor \citep{2009ApJ...702..791M} and Neil Gehrels Swift
Observatory \citep{Gehrels_2004} to benefit most from early warning detection with poor
localizations. Indeed, new observational modes enabled with the Murchinson Wide
Field Array will allow tests of BNSs as possible progenitors of non-repeating
fast radio bursts~\citep{james2019}, and will probe longstanding predictions
tying fast radio bursts to pre-merger magnetosphere
interactions~\citep{Hansen:2000am}.

Swift's ability to rapidly localize will facilitate observations of near-merger
X-ray emission, and recent Swift/BAT updates~\citep{Tohuvavohu:2020stm} will
enable subthreshold gamma-ray burst detections. Prompt X-ray observations could
help reveal the immediate aftermath of the coalescing objects, and subthreshold
detections could identify off-axis gamma-ray bursts and help probe the jet
structure associated with these mergers. Similarly, observations by the Fermi
Large Area Telescope would help complete our understanding of high-energy
gamma-ray emission; in fact the telescope had powered down just before
GRB170817A, which occluded measurements at energies $\gtrsim 100$
MeV~\citep{Kocevski:2017liw}.

By O5, early warning detections will become more common with up to $\sim 10$
detections made $10\,{\rm s}$ before merger per year. Depending on network
configuration and duty cycle, we expect that up to 80\% of these will be
localized to $\lesssim 10 \, \mathrm{deg}^2$, making them a prime target for
optical facilities that cover $\mathcal{O}(1\, \textsc{-}\, 10) \,
\mathrm{deg}^2$, such as the Zwicky Transient Facility~\citep{Graham:2019qsw},
Dark Energy Camera~\citep{Flaugher_2015}, and the highly anticipated Vera C.
Rubin Observatory~\citep{LSST:2008ijt}. Although previous optical observations
were able to capture the kilonova associated with GW170817, it was already
fading and clouded a complete understanding of the nature of the blue
ejecta~\citep{Nicholl:2017ahq,Cowperthwaite:2017dyu}. ~\cite{Chase:2021ood}
provides an in depth review of kilonova detectability prospects across multiple
observing bands for a selection of current and planned wide field-of-view
observatories.

% Comparison to Surabhi's paper

Other works have also estimated sensitivities and localization prospects for
early warning detection of BNS for specific sensitivities and 100\% duty
cycles. We find that when our SNR detection threshold is modified to match
those works, we obtain similar results in the 100\% duty cycle limit.
\cite{Sachdev:2020lfd} considered an HLV network operating at design
sensitivity, finding that $\mathcal{O}(0.1 - 1)\%$ of all detected BNS events
will be detected early with localizations $\lesssim 100\mathrm{deg}^2$. If we
impose a detection threshold of SNR 10, corresponding to the top 99\% of recovered
events in their study, we obtain similar expected localizations.
\cite{Nitz:2020vym} also considered HLV, HKLV, and AHKLV networks from the
``design'' to ``Voyager'' eras of ground based detectors. We find that the distributions
we present in Figure~\ref{fig:cdf} for the O5HLV and O5AHLV networks are
consistent with the distributions they find at SNR 10 in their Figure 3. This
is complicated by the fact that we use slightly different sensitivity curves.

% Comparison to LVK
The observing scenarios document most recently produced by the
LVK~\citep{KAGRA:2013rdx} does not consider early warning prospects, but we can
compare our $0\,{\rm s}$ early prospects to theirs. Their predictions for O4
considered the same HKLV network at 70\% duty cycle with one major
difference: KAGRA was assumed to reach 80 Mpc sensitivity. $\sim 40\%$
($\sim14\%$) of detected events were predicted to have 90\%-credible sky
areas smaller than 20 (5) $\mathrm{deg}^2$. We find that this matches our
predictions in the bottom right panel of Figure~\ref{fig:cdf} which assumes a
detection threshold of SNR 15 and a duty cycle of 70\%. If we drop our SNR threshold to 12 to
match the observing scenario document, we find slightly poorer constraints,
likely attributable to the less sensitive KAGRA used in our network.

Although our results agree with similar studies, we caution that the specifics
of the predicted distributions are highly dependent on the SNR threshold used
for recovery. As shown in Section~\ref{sec:comparison}, we expect this method
to consistently agree to within a factor of a few at the 50\% level though
Figure~\ref{fig:pp} shows that there is bias in the predicted distribution.
Our detection threshold of 15 ensures that we 1) conservatively estimate the
detection rate and 2) obtain reasonably accurate localization distributions. We
have limited our study to the current generation of ground-based detectors, but
others have considered early warning prospects for networks that include Cosmic
Explorer and the Einstein Telescope~\citep{Akcay:2018aqh,Chan:2018csa,Nitz:2021pbr}. We
leave further studies of these configurations to future work.

\section{Acknowledgments}

LIGO was constructed by the California Institute of Technology and Massachusetts
Institute of Technology with funding from the National Science Foundation and
operates under cooperative agreement PHY-1764464. This paper carries LIGO
Document Number LIGO-P2200010. The authors are grateful for computational
resources provided by the LIGO Laboratory and supported by National Science
Foundation Grants PHY-0757058 and PHY-0823459. SB further acknowledges support
from the Deutsche Forschungsgemeinschaft (DFG), project MEMI number BE6301/2-1,
and NSF grant PHY-1836779. We thank Surabhi Sachdev for providing a careful
review of this manuscript, and BS Sathyaprakash for useful comments. RM
gratefully acknowledges productive conversations with Shreya Anand and Derek Davis.

\bibliography{ms}

\end{document}